\documentclass{article}
\usepackage{spconf,amsmath,amssymb,graphicx,hyperref,makecell,cite}
\usepackage{etoolbox}

\apptocmd{\thebibliography}
  {\setlength{\itemsep}{0pt}}
  {}{}

\title{Perceptual Differences and Preferences in Speech Enhancement}
\name{Simon Dahl Jepsen \& Jesper Rindom Jensen \thanks{\begin{minipage}{0.18\linewidth} 
\hspace{-0.75cm}\includegraphics[width=1.4\linewidth]{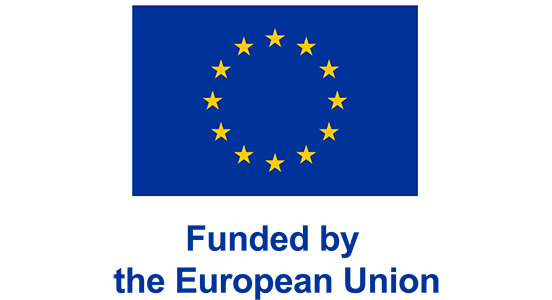}
\end{minipage}
\hfill
\hspace{-0.4cm}\begin{minipage}{0.8\linewidth} Funded by the European Union. Views and opinions expressed are however those of the author(s) only and do not necessarily reflect those of the European Union or the European Research Executive Agency. Neither the European Union nor the granting authority can be held responsible for them.
\end{minipage}}}
\address{Audio Analysis Lab, Dept. Electronic Systems, Aalborg University }
\begin{document}
\ninept

\maketitle
\raggedbottom
\begin{abstract}
Speech enhancement (SE) systems are commonly evaluated with objective metrics, with and without references, and with categorical subjective listening tests, all measuring speech quality and intelligibility. So far, it is unclear from such evaluations whether the processed speech is perceptually different or is preferred by everyday listeners. We investigate discrimination and preference as complementary dimensions of SE evaluation through crowdsourced triangle and pairwise preference tests across four SE systems, with comparisons to clean references. Results show that both listeners' ability to distinguish and whether they exhibit systematic preference vary across systems and substantially across acoustic conditions. Pairwise evaluations of objective metrics showed variable correlation with both discriminability and preference, motivating further studies into these novel aspects in speech enhancement evaluation.
\end{abstract}
\section{Introduction}
\label{sec:intro}

Distinguishing speech from noise in recordings from authentic environments is a fundamental problem in spoken language processing. This challenging inverse problem has long-standing traditions in statistical signal processing, estimating speech and noise model parameters to infer clean speech from noisy observations \cite{ephraim1984speech,gerkmann2011unbiased}. The paradigm has now transitioned to learning-based studies with multiple underlying methods, which are generally categorised into discriminative \cite{lu2025explicit,kuhne2026mambattention} and generative \cite{lemercier2023storm,li2026meanflowse} approaches. However, the goal remains the same: infer clean speech from noisy observations. 


The \textit{''gold standard''} for evaluating perceptual quality of speech enhancement is human listening tests, with established ITU-T p.800 methods including absolute (ACR), degradation (DCR), and comparative (CCR) categorical quality assessment \cite{itu_p800_2001,itu_p835_2003,itu_p808_2018}. ACR and DCR are rated on a five-point scale. CCR is rated on a seven-point scale. Results are reported as mean opinion-score (MOS, DMOS, and CMOS, respectively) \cite{itu_p800_2001}. ITU-T P. 808 specifies how to crowdsource these experiments \cite{itu_p808_2018}. 
The ITU-T P.835 standard prescribes a method for conducting subjective evaluation of speech enhancement methods, using ACR; it asks listeners to rate the enhanced speech, attending to the speech signal, background, and overall quality, respectively \cite{itu_p835_2003}. Because subjective evaluation is expensive, SE benchmarks also rely heavily on objective measures of signal fidelity, intelligibility, and perceptual quality. Popular examples include SI-SDR \cite{le2019sdr}, STOI \cite{taal2011algorithm} and ESTOI \cite{jensen2016stoi}, composite predictors of the P.835 quality dimensions \cite{hu2008evaluation}, and PESQ \cite{rix2001perceptual}. These measures are all intrusive, requiring access to a clean reference signal, which is generally unavailable for real-world recordings. This has motivated the development of non-intrusive speech-quality measures \cite{saeki2022utmos,reddy2021dnsmos,reddy2022dnsmos,mittag2021nisqa,stahl2025distillation,ragano2024scoreq}, which estimate perceptual quality directly from the observed signal. In parallel, reference-free models such as SQUIM \cite{kumar2023torchaudio} estimate established objective measures, including STOI, PESQ, and SI-SDR, without access to the corresponding clean speech signal.

SE benchmarks are commonly summarised by corpus-level means or medians of objective or subjective scores, with differences between systems interpreted as evidence of improved performance. However, differences on these scales show separation according to that measure, but do not necessarily mean listeners can distinguish the related signals. Furtermore, it is unclear if metric separation or distinguishability corresponds to a systematic preference from everyday listeners. These distinction become important as high-performing SE systems achieve increasingly similar scores on selected benchmarks. The URGENT challenges \cite{li2026icassp} results have highlighted discrepancies between objective and subjective SE evaluation \cite{li2026icassp}, while recent work has moved toward comparative evaluation, including pairwise objective quality assessment \cite{wang2023mospc,wang2026urgentmos}. ITU-T P.800 and P.808 already describe pairwise evaluation with CCR, in which listeners jointly indicate the direction and magnitude of a perceived quality difference on a seven-point scale \cite{itu_p800_2001,itu_p808_2018}. However, the magnitude of such differences may become small for high-performing SE systems, as recent CCR evaluations also observe \cite{li2026icassp}. 

This work introduces discriminability and preference as complementary dimensions to SE evaluation. First, an evaluation framework is adopted that evaluates these new dimensions through subjective listening tests. Second, a crowdsourced evaluation of four high-performing contemporary SE systems is conducted, characterising discrimination and preference across system-pairs and speech material. Finally, it is investigated whether difference scores of a range of objective metrics can predict these new SE dimensions. The web-based listening-test framework and experimental material are made publicly available to facilitate reproducible evaluation of future SE systems\footnote{Code published upon acceptance.}. 

\section{Methods}

This study enforces a paired comparison framework in contrast to most current SE evaluations. This means two systems are compared using utterance-wise metric differences as, 
\begin{equation*}
    \Delta m_{u,A,B} = m(x_A)-m(x_B),
\end{equation*}
where \(m\) is the score of an objective metric, \(u\) is the utterance index, and \(A, B\) denote the compared SE systems. This approach aligns with recent efforts \cite{wang2025urgent} and allows comparison with paired subjective evaluation. The magnitude of difference is here described as the absolute metric difference
\begin{equation}
    D^{(m)}_{u,A,B} = |\Delta m_{u,A,B}| = |m(x_A)-m(x_B)|.
\end{equation}

\subsection{Discrimination Testing}
Detecting whether the same utterance processed by different SE algorithms is perceptually different is infeasible using the cited ITU-T standards, since the experimental variable (speech quality) differs fundamentally. It is therefore proposed to adopt a method from discrimination testing in sensory evaluation. The ISO 4120 specifies the procedure for the triangle test of perceptual difference \cite{iso9001_2015}. 
Participants are presented with three stimuli of the same utterance processed by two different SE systems \(A,B\). Two stimuli are identical; one differs (positions are randomised across six possible combinations, e.g. \(A,A,B\)). The listener should identify the odd one out, making an arbitrary guess if uncertain.
\begin{equation*}
Y^{D}_{i,u,A,B}  =
\begin{cases}
1, & \text{correct},\\
0, & \text{incorrect},
\end{cases}
\Rightarrow Y^{D}_{i,u,A,B} \sim \text{Bernoulli}(p^{D}_{u,A,B}),
\end{equation*}
where \(i\) is the listener index, \(u\) is the utterance index, with chance probability \(p_0=\frac{1}{3}\). Repeated judgements follow a binomial distribution, \(k_{u,A,B}\sim \text{Binomial}(n_{u,A,B},p_{u,A,B}).\)

Statistical significance is determined according to the standard \cite{iso9001_2015}, using an exact one-sided binomial test of the alternative hypothesis, \(H_A: p_A > \frac{1}{3}\). 
A secondary analysis investigates whether the paired objective metrics predict the discriminability rate from the listening test. Objective metrics provide scalar estimates of quality and intelligibility, whereas the triangle test measures perceptual discrimination; to test whether pairwise metric separation is associated with perceptual separation, the Spearman Rank Correlation Coefficient (SRCC) is calculated from the estimate of discrimination rate: 
\begin{equation}\label{eq:disc_utt}
    \hat{p}^{D}_{u,A,B} = \frac{k_{u,A,B}}{n_{u,A,B}}, 
\end{equation}
and the correlation is then \(\rho_S(\hat{p}^{D},D^{(m)})\).
System-pair discrimination is summarised by the unweighted utterance-level estimates,
\begin{equation}\label{eq:disc_sys}
    \overline{p}^D_{A,B} = \frac{1}{U_{A,B}}\sum_{u=1}^{U_{A,B}} \hat{p}^{D}_{u,A,B}
\end{equation}
with \(95\%\) confidence interval (CI) obtained by bootstrap resampling over utterances.
The relationship between the objective metric difference and the estimated discriminability rate is modelled by a psychometric function adopted from \cite{wichmann2001psychometric}, 
\begin{equation}\label{eq:psyc}
    P(Y^{D}=1|D^{(m)})=\frac{1}{3} + \frac{2}{3}\sigma(\alpha_m+D^{(m)}\beta_m).
\end{equation}
The asymptote at \(\frac{1}{3}\) is given by the chance rate of the triangle test, the function \(\sigma(\cdot)\) is the logistic function, the parameters \(\alpha_m\) and \(\beta_m\) are estimated by maximum likelihood, and represent the intercept and slope of the curve. 

\subsection{Preference Testing}
Two stimuli (\(A, B\)) of the same utterance processed by different SE systems are presented to a listener. The listener is asked ''\textit{Listen to both recordings, then choose which you would prefer for everyday listening. Select “No preference” when neither is preferred}''. This is the paired comparison paradigm commonly referred to as \(\text{A/B/N}\) \cite{punch2001paired}. The inclusion of \textit{no preference} is in opposition to ISO 5495 \cite{iso9001_2005} for \(\text{A/B}\), and its effects are still a debated research question \cite{perez2017practical}. However, the ''\textbf{tie}'' option tends to yield higher precision \cite{punch2001paired,leijon2019bayesian}. The ''\textbf{no preference}'' category allows perceptually similar conditions to be treated as preference-equivalent and thus distinguished from genuine disagreement. 
To this end, decisiveness is defined from the response outcomes \(R\in\{A,B,N\}\) as \(Y^{\mathrm{dec}}_{u,A,B}=\mathbb{I}\{R\in \{A,B\}\}\). Decisiveness is estimated empirically from the relation
\begin{equation}\label{eq:emp_dec}
    \hat{p}^{\mathrm{dec}}_{u,A,B}=\frac{N_A+N_B}{N_A+N_B+N_N},
\end{equation}
and estimates the probability that listeners express a preference for either system under evaluation. 
The SRCC between the discrimination rate and decisiveness is calculated as \(\rho_S(\hat{p}^D,\hat{p}^{dec})\), since it is expected that no preference is expressed when two signals are indistinguishable. 
To evaluate the direction of preference, a signed preference variable is defined as:
\begin{equation}
    Y^{\mathrm{pref}}_{u,A,B} =
\begin{cases}
+1, & R=A,\\
0,  & R=N,\\
-1, & R=B.
\end{cases}
\end{equation}
The mean preference direction is then 
\(S_{A,B}=\mathbb{E}[Y^{\mathrm{pref}}]=P(R=A)-P(R=B)\),
and is estimated empirically from 
\begin{equation*}\label{eq:emp_pref}
    \hat{S}_{A,B}=\frac{N_A-N_B}{N_A+N_B+N_N},\qquad \hat{S}_{A,B} \in [-1;1],
\end{equation*}
where positive and negative scores indicate a preference for system A or B, respectively. The magnitude of this score represents both the agreement and decisiveness frequency.
Both decisiveness and preference direction are modelled as functions of \(D^{(m)}\) and \(\Delta m\) respectively. Decisiveness is modelled similarly to discrimination, but without the chance floor, and preference is modelled as a bidirectional logistic function,
\begin{align}
    P(Y^{dec}|D^{(m)})&=\sigma(\alpha^{dec}_{m}+\beta^{dec}_mD^{(m)}), \\
    \mathbb{E}\left[Y^\mathrm{pref}|\Delta{m}\right] &= 2 \sigma(\alpha_m^\mathrm{pref}+ \beta_m^\mathrm{pref}\Delta{m})-1.
    \label{eq:bidir}
\end{align}

\section{Experimental Setup}
The proposed methods are used to evaluate four state-of-the-art SE systems: the discriminative models MP-SEnet \footnote{https://github.com/yxlu-0102/MP-SENet.git} (MP)\cite{lu2025explicit} and MambAttention \footnote{https://github.com/NikolaiKyhne/MambAttention.git} (MA)\cite{kuhne2026mambattention}; the generative model MeanFlowSE \footnote{https://github.com/liduojia1/MeanFlowSE.git} (MF)\cite{li2026meanflowse}; and the hybrid StoRM \footnote{https://github.com/sp-uhh/storm.git} (Sto)\cite{lemercier2023storm}. All models are trained on the VoiceBank+DEMAND dataset \cite{valentini2017noisy}, and inference is performed using publicly available checkpoints.
The trained models enhance a selection of 45 utterances randomly sampled from the DNS2020 test set \cite{reddy2020interspeech}. Each utterance is truncated to 4 seconds, and the model outputs are loudness normalised\footnote{https://github.com/csteinmetz1/pyloudnorm}.
For each of the four chosen SE systems, a wide selection of both intrusive and non-intrusive metrics is calculated \cite{rix2001perceptual,jensen2016stoi,le2019sdr,hu2008evaluation,mittag2021nisqa,reddy2022dnsmos,stahl2025distillation,ragano2024scoreq,kumar2023torchaudio,saeki2022utmos}.
Using an incomplete pairwise design, all four systems are paired with each other and the clean references of the DNS dataset, giving 10 total system-pairs. For each \textbf{utterance}, 3 unique \textbf{system-pairs} are evaluated, while balancing the occurrence of each system and system-pair across the experiment, giving 135 utterance/system-pair \textbf{conditions}. 
Each participant completes 30 trials in both the discrimination and preference blocks. Every trial evaluates a single utterance/pair condition. A schedule is derived for each session such that the same utterance is only presented once per block and, when feasible, the same utterance/system-pair is not presented to a participant in both blocks, to minimise familiarisation effects. 
The presentation order in the triangle trials is randomised and balanced to minimise the effects of systematic guessing patterns. The presentation order of the blocks is randomised to balance the familiarisation and fatigue effects. 92 subjects without any known hearing impairment volunteered to participate anonymously. A headphone screening was performed; failing participants were excluded and are not included in the total number of subjects. 

\section{Results}

\subsection{Triangle Test}
\begin{figure}
    \centering
    \includegraphics[]{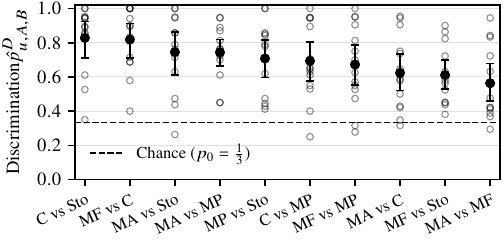}
    \vspace{-10pt}
    \caption{Triangle test discrimination rate across system pairs. Each open dot is the empirically estimated discrimination rate of an utterance/system-pair condition as in \ref{eq:disc_utt}. The filled dots represent the average across utterances as in \ref{eq:disc_sys}. The error bars represent the \(95\% \) CI by bootstrap resampling.}
    \label{fig:disc_sys_pair}\vspace{-12pt}
\end{figure}
The triangle test results show that listeners reliably distinguished the processed samples from one another and from the clean references. The estimated discrimination rate across all responses was \(\hat{p}^D = \frac{k}{n}=\frac{1775}{2541}\approx 0.7\) which is significantly above chance. 
Fig. \ref{fig:disc_sys_pair} shows that discriminability varies across system pairs, where the mean utterance-level discrimination rate (\(\overline{p}^D_{A,B}\)) is also above chance for every system pair, with the StoRM (Sto) model vs clean references being the most discriminable, with \(\overline{p}^D_{\textrm{Sto,cle}}=0.83\). Conversely, the MambAttention vs MeanFlow models have the lowest mean utterance-level discrimination rate at \(\overline{p}^D_{\textrm{MA,MF}}=0.56\). However, inter-pair comparability is limited in this experimental setup because each system pair is evaluated on different utterances. 

\begin{figure}
    \centering
    \includegraphics[]{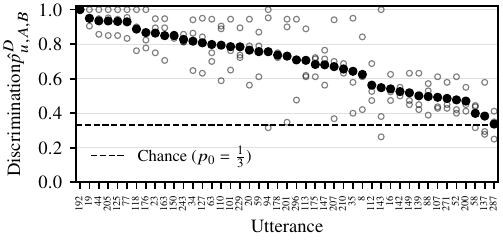}
    \vspace{-10pt}
    \caption{Triangle test discrimination rate across utterances. Each open dot is the empirically estimated discrimination rate of an utterance/system-pair condition as in \ref{eq:disc_utt}. The filled dots represent the average across system pairs for the given utterance.}
    \label{fig:disc_utt}
\end{figure}

\begin{figure}
    \centering
    \includegraphics[]{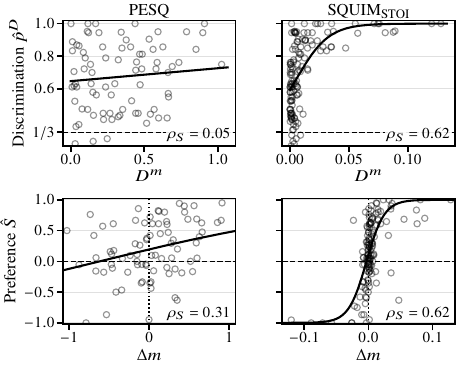}
    \vspace{-10pt}
    \caption{Empirical discrimination rate (\ref{eq:disc_sys}) and preference direction (\ref{eq:emp_pref}) a function of paired difference of PESQ and \(\mathrm{SQUIM}_\mathrm{STOI}\). The curves in the top row show the chance constrained psychometric fit (\ref{eq:psyc}), and the dashed line denotes the chance performance. The bottom row shows the bidirectional logistic fit (\ref{eq:bidir}) Conditions including clean references are excluded for PESQ.}
    \label{fig:psyc_disc}
\end{figure}

This point is further emphasised by evaluating the utterance-level discrimination rate, which is shown in Figure \ref{fig:disc_utt}. Considerable variation is observed across utterances, with mean discrimination across system pairs ranging from 0.34 to 1.0. This suggests that perceptual differences between SE systems depend on the properties of the acoustic evidence, which the triangle test itself is not able to identify. 
Although some of the variation is expected from the measurement noise introduced by the finite number of evaluations per condition, these results motivate the inclusion of utterance-wise paired evaluation, also when considering objective metrics, since system-level averages could hide perceptually important differences for individual speech signals. 
The absolute metric differences \(D^{(m)}_{u,A,B}\) are evaluated for all selected utterances, and the SRCC is calculated relative to the empirical discrimination rate of those utterances. The results are shown in Table \ref{tab:metric_srcc}. The correlation varies considerably across metrics, with the Torchaudio-Squim predictor of the STOI metric showing the strongest correlation with discrimination (\(\rho_S=0.62\)); other non-intrusive methods such as Score-Q and DistillMOS also show comparable correlation (0.59 and 0.51, respectively). Despite its widespread use in speech-enhancement benchmarking, PESQ showed almost no correlation with perceptual discriminability (\(\rho_S=0.05\)). The chance-constrained psychometric model (\ref{eq:psyc}) is fitted with maximum likelihood, as shown in Fig. \ref{fig:psyc_disc} for PESQ and \(\mathrm{SQUIM}_\mathrm{STOI}\). These results indicate that perceptual difference is an underrepresented factor in SE evaluation.

\begin{table}[t]
\centering
\vspace{-12pt}
\setlength{\tabcolsep}{2.2pt}
\caption{Spearman rank correlations between objective metric differences and perceptual responses. $\rho_D$ relates absolute metric differences
$|\Delta m|$ to triangle-test discrimination, $\rho_{\mathrm{dec}}$ relates $|\Delta m|$ to preference decisiveness, and $\rho_{\mathrm{pref}}$ relates signed metric differences $\Delta m$ to mean preference direction. Conditions containing the clean reference are excluded for intrusive metrics
($N=81$); otherwise $N=135$.}
\label{tab:metric_srcc}

\begin{tabular}{lccc|| lccc}
\hline
Metric & $\rho_D$ & $\rho_{\mathrm{dec}}$ & $\rho_{\mathrm{pref}}$
&
Metric & $\rho_D$ & $\rho_{\mathrm{dec}}$ & $\rho_{\mathrm{pref}}$ \\
\hline

$\textrm{SQUIM}_{\textrm{STOI}}$
& \textbf{0.62} & \textbf{0.56} & 0.62
&
NISQA-MOS
& 0.32 & 0.20 & 0.33 \\

SCOREQ
& 0.59 & 0.54 & \textbf{0.75}
&
$\textrm{DNSMOS}_{\textrm{OVRL}}$
& 0.29 & 0.27 & 0.31 \\

$\textrm{SQUIM}_{\textrm{PESQ}}$
& 0.57 & 0.53 & 0.62
&
CSIG
& 0.20 & 0.12 & 0.05 \\

Distill-MOS
& 0.51 & 0.51 & 0.41
&
$\textrm{DNSMOS}_{\textrm{SIG}}$
& 0.18 & 0.18 & 0.08 \\

$\textrm{SQUIM}_{\textrm{SI-SDR}}$
& 0.49 & 0.55 & 0.66
&
NISQA-Disc.
& 0.21 & 0.13 & 0.30 \\

ESTOI
& 0.43 & 0.38 & 0.48
&
COVL
& 0.17 & 0.13 & 0.17 \\

NISQA-Col.
& 0.34 & 0.24 & 0.34
&
CBAK
& 0.17 & 0.27 & 0.36 \\

$\textrm{DNSMOS}_{\textrm{BAK}}$
& 0.33 & 0.35 & 0.40
&
SI-SDR
& 0.17 & 0.18 & 0.47 \\

NISQA-Loud.
& 0.33 & 0.24 & 0.38
&
UTMOS
& 0.13 & 0.12 & 0.21 \\

NISQA-Noi.
& 0.33 & 0.22 & 0.42
&
PESQ
& 0.05 & 0.15 & 0.31 \\

\hline
\end{tabular}
\vspace{-12pt}
\end{table}

\subsection{Preference and Decisiveness}
\begin{figure}[h]
    \centering
    \includegraphics[]{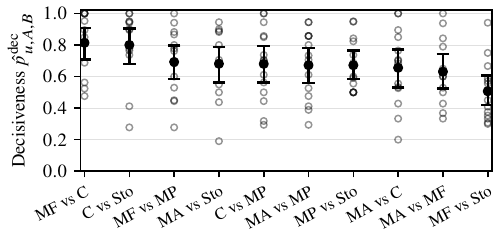}
    \vspace{-10pt}
    \caption{Decisiveness of the preference test across system pairs. Each open dot is the empirically estimated decisiveness rate of an utterance/system-pair condition as in \ref{eq:emp_dec}. The filled dots represent the average across utterances. The error bars represent the \(95\% \) CI by bootstrap resampling.}
    \label{fig:decisiveness}\vspace{-12pt}
\end{figure}
\begin{figure}
    \centering
    \includegraphics[]{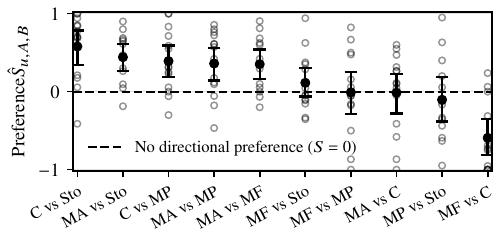}
    \vspace{-10pt}
    \caption{Directional preference across system pairs. Each open dot is the empirically estimated decisiveness rate of an utterance/system-pair condition as in \ref{eq:emp_pref}. The filled dots represent the average across utterances. The error bars represent the \(95\% \) CI by bootstrap resampling.}
    \vspace{-12pt}
    \label{fig:pref_dir}
\end{figure}

To analyse the preference experiment, an account is made towards the inclusion of a \textit{No Preference} option. Responses were analysed in terms of decisiveness, $\hat p^{\mathrm{dec}}_{u,A,B}$, describing the proportion of trials in which listeners expressed a preference for either system, and preference direction, $\hat S_{u,A,B}$, describing which system was preferred, with no-preference responses as zero-valued observations.
Decisiveness differs from discrimination, because two processed signals can be perceptually different without listeners having a preference. Across all trials, listeners expressed a preference for either system in \(68\%\) of the trials. Slightly less than the global discriminability. 
The decisiveness is evaluated empirically across system pairs as the utterance-level averages of equation \ref{eq:emp_dec}. Fig. \ref{fig:decisiveness} shows the results. Decisiveness also varies across system pairs: MF vs clean reference shows the highest average decisiveness, followed by StoRM vs clean, with mean utterance-level decisiveness rates of 0.84 and 0.79, respectively. MeanFlow vs StoRM is the least decisive pair, with a rate of 0.51. All system pairs show considerable utterance-level variation, indicating that decisiveness depends on the acoustic evidence. These results are interpreted similarly to the discrimination results in Fig. \ref{fig:disc_sys_pair}.
The \textbf{direction} of the decisive preferences is considered in Fig. \ref{fig:pref_dir}. Directional preference is observed when comparing the MF, Sto, and MP models with the MA models and clean references, with preference toward the latter. Between MF, Sto, and MP, no clear preference is observed. Interestingly, on average, the MambAttention model is equally preferred to the clean references. The results show that pairs where listeners generally express a preference (high decisiveness) do not necessarily correspond to a consistent preference direction, which motivates an analysis of the correlation with objective speech metrics. Table \ref{tab:metric_srcc} shows the SRCC between objective metrics and decisiveness and preference, respectively. As with discrimination, Score-Q and TorchAudio-Squim show the highest correlation, whereas popular metrics such as PESQ, SI-SDR, and DNSMOS show weaker correlations.

\subsection{Limitations}
The study has shown that discriminability and preference are utterance-dependent; however, some of the observed variations might also be attributed to other factors. As mentioned, a limited number of evaluations per trial can introduce variation due to measurement noise. Measurement noise could here originate from uncontrollable parameters inherent in the crowdsourcing approach. Different playback devices, individual sound pressure, listener attention, miss-hits, background noise, and language proficiency/background are unmeasured variations in this study that could introduce uncontrolled variability. Because the evaluated samples in this experiment come from a different corpus than the models were trained on, they likely introduced utterance-dependent variability due to mismatched acoustic conditions.
The choice of an incomplete pairwise experimental design allowed a broader representation of acoustic conditions, but limits inter-pair comparability, and utterance-wise averages should be compared descriptively. All reported metrics measure factors other than perceptual similarity or preference, and intuitively, two signals can be rated as similar in quality yet still be easily distinguished.

\section{Conclusion}
Speech enhancement evaluation often reduces system differences to objective or subjective quality metrics. Our results show that systems with similar objective scores can remain perceptually distinguishable, and that discriminability is not necessarily the same as having a systematic listener preference. Additionally, most objective metrics showed poor correlation with both discriminability and preference, 
Future work should focus on substantiating this study with laboratory-setting controlled listening experiments, with more repeated judgements and direct comparison with ITU standards for subjective listening tests. Probabilistic choice models could be used to account for listener and stimulus variations and provide a framework for comparing models and ensuring transitivity.

\section{Acknowledgements}
Generative AI (GPT 5.6) assisted in writing scripts and code for both experiments and plotting. Grammatical models have been used for spelling and grammar assistance. 

The authors thank Rodrigo Ordoñez and Christian Sejer Pedersen for valuable discussions about this work.

\newpage

\section{Compliance with Ethical Standards}
\vspace{-5pt}
The experimental protocol was reviewed and approved by the Aalborg University Research Ethics Committee.

\vspace{-5pt}
\begingroup
\fontsize{9}{9.5}\selectfont
\bibliographystyle{IEEEbib}
\bibliography{IEEEabrv,myabrv,refs}

@STRING{IEEE_J_ASSP       = "{IEEE} Trans. Acoust., Speech, Signal Process."}

@STRING{IEEE_J_ASL        = "{IEEE} Trans. Audio, Speech, and Language Process."}

@STRING{IEEEACM_J_ASL     = "{IEEE/ACM} Trans. Audio, Speech, and Language Process."}

@string{inter={Proc. Interspeech}}

@STRING{icassp = {Proc.\ IEEE Int.\ Conf.\ Acoust., Speech, Signal Process.}}

@STRING{jasa = {J.\ Acoust.\ Soc.\ Am.}}

@STRING{asru = {Proc.\ IEEE Automatic Speech Recognition and Understanding Workshop}}

@article{ephraim1984speech,
  title={Speech enhancement using a minimum mean-square error log-spectral amplitude estimator},
  author={Ephraim, Y. and Malah, D.},
  journal=IEEE_J_ASSP,
  volume={ASSP. 32},
  number={6},
  pages={1109--1121},
  year={1984},
  publisher={IEEE}
}

@article{gerkmann2011unbiased,
  title={Unbiased {MMSE}-based noise power estimation with low complexity and low tracking delay},
  author={Gerkmann, T. and Hendriks, R. C.},
  journal=IEEE_J_ASL,
  volume={20},
  number={4},
  pages={1383--1393},
  year={2011},
  publisher={IEEE}
}

@techreport{itu_p800_2001,
  author      = {{International Telecommunication Union}},
  title       = {Methods for Subjective Determination of Transmission Quality},
  institution = {International Telecommunication Union},
  type        = {Recommendation},
  number      = {ITU-T P.800},
  address     = {Geneva, Switzerland},
  year        = {2001}
}

@techreport{itu_p835_2003,
  author      = {{International Telecommunication Union}},
  title       = {Subjective Test Methodology for Evaluating Speech Communication Systems That Include Noise Suppression Algorithm},
  institution = {International Telecommunication Union},
  type        = {Recommendation},
  number      = {ITU-T P.835},
  address     = {Geneva, Switzerland},
  year        = {2003}
}

@techreport{itu_p808_2018,
  author      = {{International Telecommunication Union}},
  title       = {Subjective Evaluation of Speech Quality with a Crowdsourcing Approach},
  institution = {International Telecommunication Union},
  type        = {Recommendation},
  number      = {ITU-T P.808},
  address     = {Geneva, Switzerland},
  year        = {2018}
}

@inproceedings{le2019sdr,
  title={{SDR}--half-baked or well done?},
  author={Le Roux, J. and Wisdom, S. and Erdogan, H. and Hershey, J. R.},
  booktitle=icassp,
  pages={626--630},
  year={2019}
}

@inproceedings{rix2001perceptual,
  title={Perceptual evaluation of speech quality ({PESQ})-a new method for speech quality assessment of telephone networks and codecs},
  author={Rix, A. W. and Beerends, J. G. and Hollier, M. P. and Hekstra, A. P.},
  booktitle=icassp,
  volume={2},
  pages={749--752},
  year={2001}
}

@article{taal2011algorithm,
  title={An algorithm for intelligibility prediction of time--frequency weighted noisy speech},
  author={Taal, C. H. and Hendriks, R. C. and Heusdens, R. and Jensen, J.},
  journal=IEEE_J_ASL,
  volume={19},
  number={7},
  pages={2125--2136},
  year={2011},
  publisher={IEEE}
}

@article{jensen2016stoi,
  author  = {Jensen, J. and Taal, C. H.},
  title   = {An Algorithm for Predicting the Intelligibility of Speech Masked by Modulated Noise Maskers},
  journal = IEEE_J_ASL,
  volume  = {24},
  number  = {11},
  pages   = {2009--2022},
  year    = {2016}
}

@inproceedings{reddy2021dnsmos,
  title={{DNSMOS}: A non-intrusive perceptual objective speech quality metric to evaluate noise suppressors},
  author={Reddy, C. K. A. and Gopal, V. and Cutler, R.},
  booktitle=icassp,
  pages={6493--6497},
  year={2021}
}

@inproceedings{reddy2022dnsmos,
  title={{DNSMOS P. 835}: A non-intrusive perceptual objective speech quality metric to evaluate noise suppressors},
  author={Reddy, C. K. A. and Gopal, V. and Cutler, R.},
  booktitle=icassp,
  pages={886--890},
  year={2022}
}

@inproceedings{mittag2021nisqa,
  title     = {{NISQA: A Deep {CNN}-Self-Attention Model for Multidimensional Speech Quality Prediction with Crowdsourced Datasets}},
  author    = {G. Mittag and B. Naderi and A. Chehadi and S. M{\"o}ller},
  year      = {2021},
  booktitle = inter,
  pages     = {2127--2131},
  doi       = {10.21437/Interspeech.2021-299},
  issn      = {2958-1796},
}

@inproceedings{kumar2023torchaudio,
  title={Torchaudio-squim: Reference-less speech quality and intelligibility measures in torchaudio},
  author={Kumar, Anurag and Tan, Ke and Ni, Zhaoheng and Manocha, Pranay and Zhang, Xiaohui and Henderson, Ethan and Xu, Buye},
  booktitle={Proc.\ IEEE Int.\ Conf.\ Acoust., Speech, Signal Process.},
  pages={1--5},
  year={2023}
}

@article{lemercier2023storm,
  title={{StoRM}: A diffusion-based stochastic regeneration model for speech enhancement and dereverberation},
  author={Lemercier, J.-M. and Richter, J. and Welker, S. and Gerkmann, T.},
  journal=IEEEACM_J_ASL,
  volume={31},
  pages={2724--2737},
  year={2023},
  publisher={IEEE}
}

@inproceedings{li2026meanflowse,
  title={{MeanflowSE}: One-step generative speech enhancement via conditional mean flow},
  author={Li, D. and Lu, S. and Pan, H. and Zhan, Z. and Hong, Q. and Li, L.},
  booktitle=icassp,
  pages={16172--16176},
  year={2026}
}

@article{lu2025explicit,
  title={Explicit estimation of magnitude and phase spectra in parallel for high-quality speech enhancement},
  author={Lu, Y.-X. and Ai, Y. and Ling, Z.-H.},
  journal={Neural Networks},
  volume={189},
  pages={107562},
  year={2025},
  publisher={Elsevier}
}

@article{kuhne2026mambattention,
  title={{MambAttention}: {Mamba} with multi-head attention for generalizable single-channel speech enhancement},
  author={K{\"u}hne, N. L. and Jensen, J. and {\O}stergaard, J. and Tan, Z.-H.},
  journal=IEEE_J_ASL,
  year={2026},
  publisher={IEEE}
}

@misc{iso9001_2015,
  author       = {{International Organization for Standardization}},
  title        = {{Sensory Analysis - Methodology - Triangle Test}},
  howpublished = {{ISO Standard No. 4120:2021}},
  year         = {2021},
  publisher    = {ISO},
  address      = {Geneva, Switzerland},
}

@misc{iso9001_2005,
  author       = {{International Organization for Standardization}},
  title        = {{Sensory Analysis - Methodology - Paired Comparison Test}},
  howpublished = {{ISO Standard No. 5495:2005}},
  year         = {2005},
  publisher    = {ISO},
  address      = {Geneva, Switzerland},
}

@article{perez2017practical,
  title={A practical guide and software for analysing pairwise comparison experiments},
  author={Perez-Ortiz, Maria and Mantiuk, Rafal K},
  journal={arXiv preprint arXiv:1712.03686},
  year={2017}
}

@inproceedings{li2026icassp,
  title={{ICASSP} 2026 urgent speech enhancement challenge},
  author={Li, C. and Wang, W. and Sach, M. and Zhang, W. and Saijo, K. and Cornell, S. and others},
  booktitle=icassp,
  pages={21919--21921},
  year={2026}
}

@article{wang2026urgentmos,
  title={{UrgentMOS}: Unified multi-metric and preference learning for robust speech quality assessment},
  author={Wang, W. and Zhang, W. and Li, C. and Wang, J. and Cornell, S. and Sach, M. and Saijo, K. and Fu, Y. and Ni, Z. and Han, B. and others},
  journal={arXiv preprint arXiv:2601.18438},
  year={2026}
}

@inproceedings{wang2023mospc,
  title={{MOSPC}: {MOS} prediction based on pairwise comparison},
  author={Wang, K. and Zhao, Y. and Dong, Q. and Ko, T. and Wang, M.},
  booktitle={Proc.\ Annual Meeting of the Association for Computational Linguistics},
  pages={1547--1556},
  year={2023}
}

@inproceedings{wang2025urgent,
  title={Urgent-pk: Perceptually-aligned ranking model designed for speech enhancement competition},
  author={Wang, J. and Li, C. and Wang, W. and Zhang, W. and Cornell, S. and Sach, M. and Scheibler, R. and Saijo, K. and Fu, Y. and Ni, Z. and others},
  booktitle=asru,
  pages={1--7},
  year={2025}
}

@article{punch2001paired,
  title={Paired-comparison hearing aid preferences: Evaluation of an unforced-choice paradigm},
  author={Punch, J. L. and Rakerd, B. and Amlani, A. M.},
  journal={J. American Academy of Audiology},
  volume={12},
  number={04},
  pages={190--201},
  year={2001},
  publisher={Thieme Medical Publishers, Inc.}
}

@article{leijon2019bayesian,
  title={Bayesian analysis of paired-comparison sound quality ratings},
  author={Leijon, A. and Dahlquist, M. and Smeds, K.},
  journal=jasa,
  volume={146},
  number={5},
  pages={3174--3183},
  year={2019},
  publisher={AIP Publishing}
}

@dataset{valentini2017noisy,
  author    = {Valentini-Botinhao, C.},
  title     = {Noisy Speech Database for Training Speech Enhancement Algorithms and {TTS} Models},
  year      = {2017},
  note      = {University of Edinburgh, School of Informatics, Centre for Speech Technology Research (CSTR)},
  doi       = {10.7488/ds/2117},
  url       = {https://doi.org/10.7488/ds/2117}
}

@inproceedings{reddy2020interspeech,
  title={The {Interspeech} 2020 deep noise suppression challenge: Datasets, subjective testing framework, and challenge results},
  author    = {C. K. A. Reddy and V. Gopal and R. Cutler and E. Beyrami and R. Cheng and H. Dubey and others},
  year      = {2020},
  booktitle = inter,
  pages     = {2492--2496},
  doi       = {10.21437/Interspeech.2020-3038},
  issn      = {2958-1796}
}

@article{hu2008evaluation,
  title={Evaluation of objective quality measures for speech enhancement},
  author={Hu, Y. and Loizou, P. C.},
  journal=IEEEACM_J_ASL,
  volume={16},
  number={1},
  pages={229--238},
  year={2008},
  publisher={IEEE}
}

@inproceedings{saeki2022utmos,
  title     = {{UTMOS}: {UTokyo-SaruLab} System for {VoiceMOS} Challenge 2022},
  author    = {T. Saeki and D. Xin and W. Nakata and T. Koriyama and S. Takamichi and H. Saruwatari},
  year      = {2022},
  booktitle = inter,
  pages     = {4521--4525},
  doi       = {10.21437/Interspeech.2022-439},
  issn      = {2958-1796},
}

@INPROCEEDINGS{stahl2025distillation,
      title={Distillation and Pruning for Scalable Self-Supervised Representation-Based Speech Quality Assessment},
      booktitle=icassp, 
      author={B. Stahl and H. Gamper},
      year={2025},
      doi={10.1109/ICASSP49660.2025.10888007},
      eprint={2502.05356},
      archivePrefix={arXiv},
      primaryClass={eess.AS},
      url={https://arxiv.org/abs/2502.05356}, 
}

@article{ragano2024scoreq,
  title={{SCOREQ}: Speech quality assessment with contrastive regression},
  author={Ragano, A. and Skoglund, J. and Hines, A.},
  journal={Advances in Neural Information Processing Systems},
  volume={37},
  pages={105702--105729},
  year={2024}
}

@article{wichmann2001psychometric,
  title={The psychometric function: I. Fitting, sampling, and goodness of fit},
  author={Wichmann, F. A. and Hill, N. J.},
  journal={Perception \& Psychophysics},
  volume={63},
  number={8},
  pages={1293--1313},
  year={2001},
  publisher={Springer}
}
\endgroup
\end{document}